\documentclass[conference]{IEEEtran}
\usepackage{graphicx,
	psfrag,
	epsfig,
	epstopdf,
	amsthm,
	amssymb,
	url,
	subfigure,
	algorithm,
	algorithmic,
	balance,
	enumerate,
	color,
	url,
	setspace,
	tikz,
}
\usepackage[normalem]{ulem}
\usepackage{amsmath}
\usepackage[nospace,noadjust]{cite}
\usepackage{pbox}
\usepackage{geometry}
\usepackage{microtype}
\usepackage{multirow}

\newgeometry{left=1.6cm,right=1.6cm,bottom=1.6cm, top=1.6cm}

\newcommand{\ignore}[1]{}

\begin{document}
\title{Ensemble-based Feature Selection and Classification Model for DNS Typo-squatting Detection}
\author{
\IEEEauthorblockN{Abdallah Moubayed\IEEEauthorrefmark{1}, Emad Aqeeli\IEEEauthorrefmark{2}, and Abdallah Shami\IEEEauthorrefmark{1}}
		
\IEEEauthorblockA{\IEEEauthorrefmark{1} Western University, London, Ontario, Canada \\
	emails: \{amoubaye, abdallah.shami\}@uwo.ca
}
\IEEEauthorblockA{\IEEEauthorrefmark{2} Royal Commision Yanbu Colleges and Institutes, Yanbu, Kingdom of Saudi Arabia \\
	e-mail: emadaqeeli@rcyci.edu.sa	
}
}
\maketitle

\begin{abstract}
Domain Name System (DNS) plays in important role in the current IP-based Internet architecture. This is because it performs the domain name to IP resolution. However, the DNS protocol has several security vulnerabilities due to the lack of data integrity and origin authentication within it. This paper focuses on one particular security vulnerability, namely typo-squatting. Typo-squatting refers to the registration of a domain name that is extremely similar to that of an existing popular brand with the goal of redirecting users to malicious/suspicious websites. The danger of typo-squatting is that it can lead to information threat, corporate secret leakage, and can facilitate fraud. This paper builds on our previous work in \cite{Moubayed_DNS}, which only proposed majority-voting based classifier, by proposing an ensemble-based feature selection and bagging classification model to detect DNS typo-squatting attack. Experimental results show that the proposed framework achieves high accuracy and precision in identifying the malicious/suspicious typo-squatting domains (a loss of at most 1.5\% in accuracy and 5\% in precision when compared to the model that used the complete feature set) while having a lower computational complexity due to the smaller feature set (a reduction of more than 50\% in feature set size).
\end{abstract}


\section{Introduction}\label{Intro_dns}
\indent The Domain Name System (DNS) protocol is an important pillar in the Internet's current and future architecture \cite{DNS_definition,future_internet,Aqeeli1,Aqeeli2}. This is because it is the standard mechanism for name to IP address resolution \cite{DNS_definition}. Moreover, it helps users to determine the location of servers and mailing hosts, resulting in a direct impact on the data exchange process \cite{DNS_definition,future_internet}.\\
\indent However, DNS is vulnerable to a variety of security threats and attacks, as illustrated by the recent DNS attacks \cite{DNS_vulnerabilities,DNS_vulnerabilities1}. One example is the distributed denial of service (DDoS) attack on Dyn in October 2016 which resulted in a significant portion of America's Internet Service to go down \cite{DNS_attacks1,DNS_attacks2}. Another example is the attack on a Brazilian Bank's website. During this attack, attackers rerouted the traffic targeted to the bank's website to their own servers. This was done by changing the DNS registrations of all the bank's domains, resulting in many users divulging their authentication information to the malicious attackers \cite{DNS_attacks3}. These vulnerabilities can be mainly attributed to the lack of data integrity and origin authentication processes included within the DNS protocol structure.\\
\indent One such vulnerability that the DNS protocol suffers from is that of typo-squatting. Typo-squatting refers to the registration of a domain name that is extremely similar to that of an existing popular brand with the goal of redirecting users to malicious/suspicious websites. This is done by registering confusingly similar domain names that the user might not pay attention to \cite{DNS_vulnerabilities4}. For example, the \textit{www.paypal.com} domain can be easily confused with \textit{www.paypa1.com} domain. The danger of typo-squatting is that it can lead to information threat, corporate secret leakage, and can facilitate fraud \cite{DNS_vulnerabilities2,R1_1}.\\
\indent Hence, it is crucial that DNS is able to tolerate failure and is resilient to attacks given its importance to the proper functioning of the Internet \cite{DNS_vulnerabilities1}. This has led to various researchers proposing different mechanisms to combat and protect against failures and attacks. One such mechanism is the DNSSEC protocol which aims at addressing some of the security vulnerabilities of DNS by providing data integrity and origin authentication \cite{dnssec_protocol}. Yet, DNSSEC still can not address other attacks such as amplified denial of service attacks \cite{DNS_vulnerabilities1,dnssec_performance}. Thus, it is important that more efficient detection mechanisms are implemented that can protect systems from the various attacks by better identifying malicious queries. \\
\indent This paper builds on our previous work in \cite{Moubayed_DNS} which only proposed majority-voting based classifier to detect DNS typo-squatting. In contrast, this work proposes an ensemble-based feature selection and classification (EFSBC) model to detect DNS typo-squatting attack. This is done to reduce the complexity of the DNS typo-squatting detection framework while maintaining its high accuracy and low false positive rate. To that end, this work presents a framework in which three different feature selection techniques are combined to identify features that are crucial for the accurate detection of malicious/suspicious DNS domains. Moreover, the framework proposes the use of bagging ensemble classification models (that can reduce model variance) to further improve the accuracy of DNS typo-squatting detection.\\
\indent The contributions of this work can be summarized as follows:
\begin{itemize}
	\item \textit{Proposing} an ensemble feature selection method that selects the crucial features using multiple selection techniques.
	\item \textit{Proposing} a bagging ensemble classification model that identifies malicious/suspicious domain names with high accuracy.
	\item \textit{Evaluating} the performance of the proposed model in comparison to other traditional classification models. 
\end{itemize}
\begin{figure}[!h]
	\centering
	\includegraphics[scale=.35]{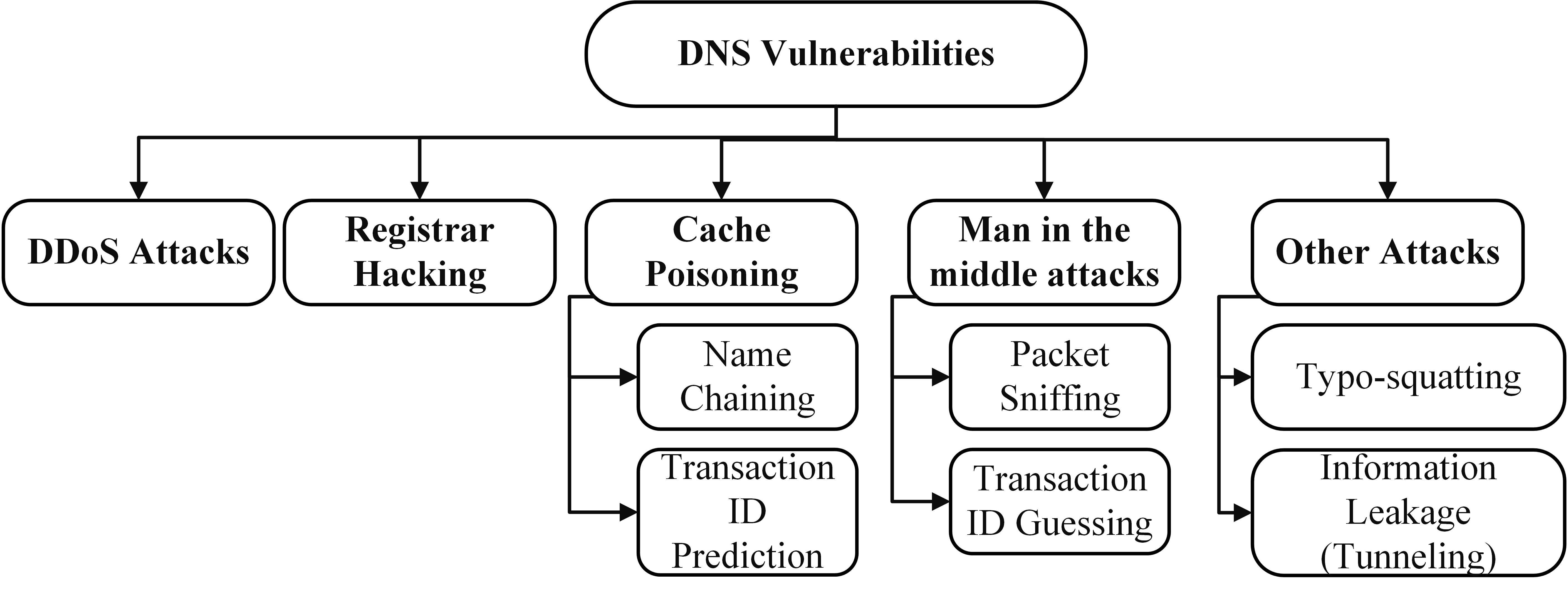}
	\caption{DNS Vulnerabilities and Challenges}
	\label{vulnerabilities}
\end{figure}
\indent The remainder of this paper is organized as follows: Section \ref{challenges_dns} describes the different security vulnerabilities that the DNS protocol faces. Section \ref{related_work_dns} summarizes the previous work in the literature. Section \ref{proposed_approach_dns} illustrates the proposed ensemble-based DNS typo-squatting detection framework and discusses its complexity. Section \ref{dataset_description_dns} describes the dataset considered in this work and the data transformation/feature extraction process. Section \ref{results_dns} presents the experiment setup and discusses the corresponding results. Finally, Section \ref{conc_dns} concludes the paper.
\section{DNS Vulnerabilities \& Challenges}\label{challenges_dns}
\indent As mentioned earlier, the DNS protocol has several security vulnerabilities due to the lack of data integrity and origin authentication within it. Fig.\ref{vulnerabilities} briefly lists some of the different vulnerabilities and attacks that DNS faces \cite{DNS_vulnerabilities1,DNS_vulnerabilities2,DNS_vulnerabilities3}.
\begin{enumerate}
\item DDoS attacks: Root DNS services are vulnerable to DDoS attacks. This is mainly due to the hierarchical architecture adopted. This is dangerous as it can cause a loss of availability of name resolution services which can lead to the stoppage of Internet service \cite{DNS_attacks1,DNS_attacks2,SDP1,SDP2,SDP3}. 
\item Registrar hijacking: Malicious users can hijack a registrar. As a result, these users would control all the corresponding domain names. In turn, this can lead to enterprises and companies losing their domain names. One such example is the attack on a Brazilian Bank's website \cite{DNS_attacks3}. As part of this attack, all the traffic to the bank's website was redirected to the attackers own servers. This was possible because the attackers changed the DNS registrations of all the bank's domains \cite{DNS_attacks3}. This attack had severe consequences with thousands of users being affected due to the leakage of sensitive information such as their banking, email, and FTP credentials \cite{DNS_attacks3}.
\item Cache Poisoning Problems: Cache poisoning is the result of the lack of data update propagation or invalidations mechanisms to DNS caches. Hence, cache poisoning can be achieved using Name Chaining or Transaction ID Prediction.
\begin{enumerate}[i-]
	\item Name Chaining: Attacker adds random DNS names in the DNS response which leads to the introduction of false information into the cache.
	\item Transaction ID Prediction: Attacker sends multiple DNS queries for domain names under his/her control. Then the attacker hopes that the transaction ID in one of the subsequent spoof replies matches the transaction ID that is used as part of the queries between the two servers. 
\end{enumerate}
\item Man in the middle (MiTM) attacks: Attacks such as Packet Sniffing and Transaction ID Guessing are possible due to the fact that the DNS protocol does not offer a mechanism for servers to provide authentication details for the data sent to clients. This can result in a threat to the users' privacy by directing them to suspicious or malicious domains and servers.
\begin{enumerate}[i-]
	\item Packet Sniffing: DNS reply packets can be intercepted and modified by the attacker.
	\item Transaction ID Guessing: Attackers that can correctly guess the transaction ID can send false replies to legitimate queries.
\end{enumerate} 
\item Other DNS attacks: In addition to the attacks listed above, DNS is also prone to other types of attacks such as Information Leakage and Typo-squatting.
\begin{enumerate}[i-]
	\item Information Leakage (DNS Tunneling): As part of this attack, an attacker would leak sensitive information as part of DNS queries or their responses.
	\item Typo-squatting: This attack focuses on registering a domain name that highly matches that of an existing domain name in an attempt to confuse/fool users. This is dangerous as it can lead to information threat, corporate secret leakage, and can facilitate fraud \cite{DNS_vulnerabilities2}.
\end{enumerate}
\end{enumerate}
This work mainly focuses on the typo-squatting attack. This is due to the severe consequences of such an attack including leakage of corporate secrets, leakage of sensitive personal information, and ultimately fraud \cite{DNS_vulnerabilities2}. Therefore, detecting such attacks through efficient and intelligent mechanisms is a necessity.
\section{Related Work}\label{related_work_dns}
\indent Securing the Internet has been a growing concern in recent years given the growth in the number of attacks witness on Internet services. Due to the abundance of data being collected by Internet service providers and network administrators, machine learning (ML)-based mechanisms have been proposed as a potential and viable efficient solution to help better detect attacks on Internet services. For example, intrusion and DDoS attack detection mechanisms using different classification algorithms such as artificial neural networks and support vector machines (SVM) have been proposed for Software-Defined Networks (SDNs) \cite{SDN_ddos_attack}. Similarly, an optimized ML-based anomaly detection framework was proposed that achieved high accuracy and low false alarm rate \cite{Injadat_BO}. Additionally, decision tree (DT)-based algorithms have also been proposed as effective DDoS attack detection mechanisms in cloud computing environments \cite{cloud_ddos_attack}. Similarly, a tree-based intrusion detection system for autonomous vehicles was proposed that achieves high detection rate with a low computational cost \cite{Li_IDS}.\\
\indent Few works in the literature explored the use of ML within the context of DNS security. Zhauniarovich \textit{et al.} surveyed the state of the art work on malicious domain detection through DNS data anaylysis \cite{R2_5}. Bilge \textit{et al.} proposed a DT-based classification model to detect malicious domains \cite{DNS_vulnerabilities3}. Similarly, Sivakorn \textit{et al.} proposed the use of ML to detect malicious DNS queries \cite{R2_1}. Also, Sountharrajan \textit{et al.} used deep learning models to detect phishing URLs \cite{R2_4}. On the other hand, Almusawi proposed an SVM model to detect DNS tunneling \cite{R2_2}.  In contrast, Fukuda \textit{et al.} proposed the use of ML to classify originator activity of DNS backscatter \cite{R2_3}. Weber \textit{et al.} proposed the use of unsupervised clustering to identify malicious domain campaigns \cite{R2_6}. However, very few focused literature works focused on the DNS typo-squatting attack.
\section{Proposed Approach}\label{proposed_approach_dns}
\subsection{Proposed Approach}
\begin{figure}[!h]
	\centering
	\includegraphics[trim=1cm 0cm 1cm 0cm,scale=.35]{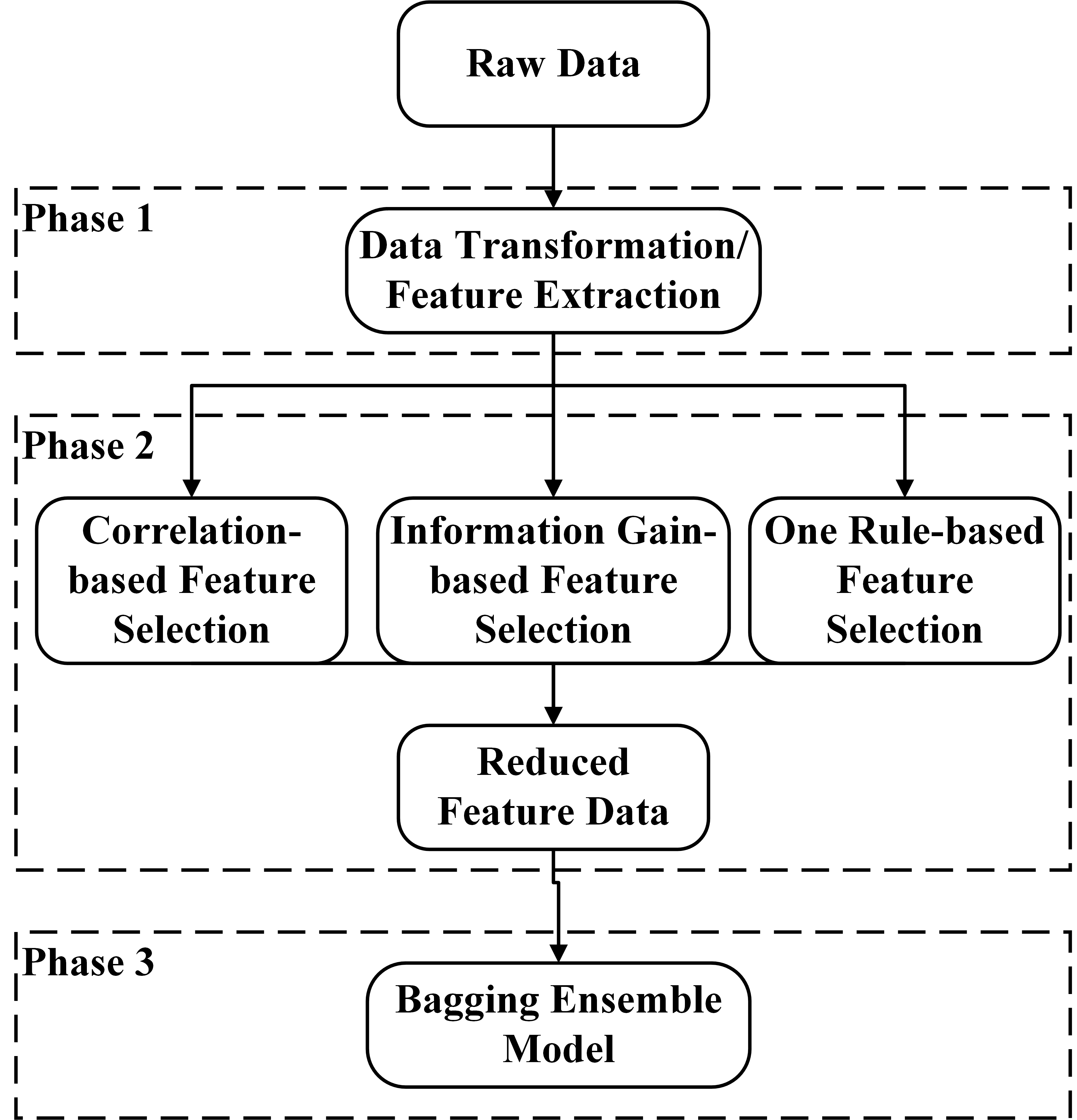}
	\caption{Proposed EFSBC Approach}
	\label{approach_fig}
\end{figure}

\indent This paper extends our previous work in \cite{Moubayed_DNS} by proposing an ensemble-based feature selection and bagging classification (EFSBC) model to detect DNS domain typo-squatting. This is done to reduce the complexity of the DNS typo-squatting detection framework while maintaining its high accuracy and low false positive rate. The proposed approach, as shown in Fig. \ref{approach_fig}, can be divided into three components, namely:
\begin{enumerate}
	\item Extract domain name representative features.
	\item Develop an ensemble feature selection model to identify crucial features. 
	\item Develop a bagging ensemble classification model to detect malicious/suspicious domains. 
\end{enumerate}
\subsection{Proposed Approach Application}
\indent As mentioned earlier, the proposed approach can be divided into three main phases, namely the data transformation/feature extraction phase, feature selection phase (using three different feature selection mechanisms), and classification model phase. In the data transformation/feature extraction phase, a set of features representative of the typo-squatting attack are extracted. These features mainly focus on the domain name and its characteristic. Section \ref{dataset_extraction} discusses this in more details.\\
\indent The second phase, a sub-set of features are selected using different feature selection mechanisms to be given to the classification model as input. This is done in an attempt to reduce the complexity of the classification model and decrease its training time without sacrificing its performance \cite{FS_reason}. This is particularly important when dealing with large scale systems generating big data \cite{FS_reason}. Three different feature selection mechanisms are considered in this work representing three different categories of feature selection algorithms. The first algorithm is the correlation-based feature selection algorithm which belongs to the group of ``Traditional Statistical'' feature selection techniques \cite{FS_CFS1}. The second algorithm is the information gain algorithm which belongs to the group of ``Information Theory'' techniques \cite{FS_IG1}. The third feature selection algorithm is the One Rule algorithm which is one of the ``Decision Tree'' based feature selection algorithms \cite{FS_OR1}. The results of these feature selection mechanisms are combined to produce a subset of features that are the highest ranked features from the dataset. The mathematical details of these algorithms are given in Section \ref{dataset_FS}.\\
\indent After performing feature selection, the reduced feature dataset is given as an input to the bagging ensemble model classifier. In particular, a bagging ensemble model is chosen as it can reduce the classification variance of any base model while maintaining their low bias characteristics and improving the classification accuracy \cite{ensemble_variance}. Therefore, a bagging ensemble classification model is adopted in this work with the aim of providing a domain typo-squatting detection framework with high accuracy and low false alarm rate.
\subsection{Complexity of Proposed Approach}
\indent The complexity of the proposed approach is dependent on the complexity of each of its phases, namely the feature extraction phase, the feature selection phase, and the classification model building phase. It is assumed that there are $M$ data samples and $N$ features. Accordingly, the complexity of the feature extraction phase is $O(M)$ as the algorithm needs to go through all the dataset to extract the $N$ features.\\ 
\indent The complexity of the feature selection phase depends on the complexity of each of the feature selection methods considered. The complexity of Correlation-based feature selection is $O(MN^2)$ to calculate all the class-feature and feature-feature correlations \cite{FS_CFS2}. On the other hand, the complexity of information gain-based feature selection method is $O(MN)$ to calculate the joint probabilities of the class-feature interaction \cite{FS_IG2}. Finally, the complexity of One Rule algorithm is $O(MN)$ given that you have to determine the classification accuracy based on each feature \cite{FS_OR_complexity}. Therefore, the overall complexity of the feature selection process is $O(MN^2)$.\\
\indent Finally, the complexity of the bagging ensemble classification model is dependent on the type of base learners used as part of the ensemble. In this work, the base learners considered are DTs and $K-$nearest neighbors ($K$NN) due to their high accuracy as shown in \cite{Moubayed_DNS}. Given that building the bagging ensemble can be performed in parallel, the complexity of the DT-based bagging ensemble can be estimated as $O(M^2N_{red})$ and the complexity of nearest neighbor-based bagging ensemble can be estimated as $O(MN_{red})$ \cite{complexity1,complexity2} where $N_{red}$ is the size of the reduced feature set.\\
\indent By combining the computational complexity of the different phases knowing that $N_{red} < N$, the overall complexity of the proposed approach is in the order of $O(MN^2+M^2N_{red})$ assuming that the DT-based bagging ensemble classification model is chosen. However, given that the feature selection process can be performed offline, the complexity of the proposed approach can be considered to be in the order of $O(M^2N_{red})$.
\section{Dataset Description}\label{dataset_description_dns}
\subsection{Data Preprocessing:}
\indent The dataset under consideration in this work was originally collected by the authors of the ``Data Driven Security'' book \cite{labeled_data_source1}. The collection process consisted of a combination of Alexa's top 1 million legitimate domains and Cryptolocker's list of domains generated algorithmically (DGA) \cite{labeled_data_source}. The resulting dataset is a list of 133,926 unique domains divided into 81,261 legitimate domains and 52,665 DGA domains. Each record consists of three fields as illustrated in Table \ref{raw_data_example}.
\begin{table}[!htbp]
	\centering
	\caption{Domain Features Description}
	\scalebox{0.75}{
		\begin{tabular}{|p{2cm}|p{3.5cm}|p{2.8cm}|}	\hline
			Field & Description & Example \\ \hline
			Host &Domain's complete url&www.mydaily.co.uk\\ \hline
			Domain&Actual domain accessed&mydaily\\ \hline
			Domain Class&Domain classification&Legit or DGA\\ \hline
		\end{tabular}
	}
	\label{raw_data_example}
\end{table}
\subsection{Data Transformation/Feature Extraction:}\label{dataset_extraction}
\indent The dataset was transformed using MATLAB into a new dataset of eight features that characterize a unique domain name. More specifically, these features were chosen due to the nature of the typo-squatting attack which mainly focuses on modifying the domain name. All the features under consideration are numeric in nature. More specifically, the first four are integers and the remaining being continuous. \\
\indent In addition to the extracted features, a binary feature representing the domain class was also added to the new dataset. In particular, a DGA domain was represented as 1 while a legitimate domain was represented as 0. Table \ref{table_of_metrics_dns} shows the value type and range of each of the aforementioned features.
\begin{table}[!h]
	\centering
	\caption{Domain Features Description}
	\scalebox{0.75}{
		\begin{tabular}{|p{5.3cm}|p{1.3cm}|p{1.5cm}|}	\hline
			Feature & Value Type & Range of Values \\ \hline
			Length of Domain Name&Numeric&[1,2,...,68]\\ \hline
			Number of Unique Characters&Numeric&[1,2,...,36] \\ \hline
			Number of Unique Letters&Numeric&[1,2,...,26]\\ \hline
			Number of Unique Numbers&Numeric&[0,1,...,10]\\ \hline
			Ratio of Letters to Domain Length&Numeric&[0-1]\\ \hline	
			Ratio of Numbers to Domain Length&Numeric&[0-1]\\ \hline
			Ratio of Unique Letters to Unique Characters&Numeric&[0-1]\\ \hline
			Ratio of Unique Numbers to Unique Characters&Numeric&[0-1]\\ \hline
			Domain Class&Numeric&[0,1]\\ \hline
		\end{tabular}
	}
	\label{table_of_metrics_dns}
\end{table}
\subsection{Feature Selection Techniques' Background:}\label{dataset_FS}
\subsubsection{Correlation-based Feature Selection}\mbox{}\\
\indent Correlation-based feature selection (CFS) is a simple algorithm that selects feature subsets based on their correlation with the class to be predicted \cite{FS_CFS2}. In essence, CFS consider a feature to be relevant if it is correlated with or predictive of the class \cite{FS_CFS2,FS_CFS3}. CFS mainly uses Pearson's correlation coefficient as its feature subset evaluation function. Accordingly, the evaluation function is \cite{FS_CFS2}:
\begin{equation}
	M_S=\frac{k \times \overline{r_{cf}}}{\sqrt{k + k\times(k-1)\times\overline{r_{ff}}}}
\end{equation}
where:
\begin{itemize}
	\item $M_S$: merit of the feature subset $S$
	\item $k$: number of features in feature subset $S$
	\item $\overline{r_{cf}}$: average class-feature Pearson correlation
	\item $\overline{r_{cf}}$: average feature-feature Pearson correlation
\end{itemize}
Using this equation, the feature subsets can be ranked and the subset with the highest correlation with the class to be predicted can be selected.
\subsubsection{Information Gain-based Feature Selection}\mbox{}\\
\indent Information gain-based feature selection is based on the use of information theory concepts such as entropy and mutual information \cite{FS_IG2}. This algorithm selects features based on the amount of information (in bits) that can be gained from these features. Accordingly, the feature evaluation function is \cite{FS_IG2}:
\begin{equation}
\begin{split}
I(F;C) & = H(F)-H(F|C) \\
       & = \sum\limits_{f_i \in F}\sum\limits_{c_j \in C} P(f_i,c_j)log\frac{P(f_i,c_j)}{P(f_i)\times P(c_j)}
\end{split}
\end{equation}
where:
\begin{itemize}
	\item $I(F;C)$: mutual information between feature subset $F$ and class $C$
	\item $H(F)$: entropy/uncertainty of discrete feature subset $F$
	\item $H(F|C)$: conditional entropy/uncertainty of discrete feature subset $F$ given class $C$
	\item $P(f_i,c_j)$: joint probability of feature having a value $f_i$ and class being $c_j$
	\item $P(f_i)$: probability of feature having a value $f_i$
	\item $P(c_j)$:probability of class being $c_j$
\end{itemize}
Using these values, the information gained from each feature with respect to the class can be calculated and the highest features can be selected.
\subsubsection{One Rule-based Feature Selection}\mbox{}\\
\indent One rule, also commonly referred to as ``OneR'' or ``1R'', algorithm is a simple one-level decision tree algorithm that creates one rule for each feature in the training data and provides an accuracy measure for that  feature \cite{FS_OR2}. The main motivation is that such a feature selection algorithm can achieve high accuracy while still providing simple rules for humans to interpret and understand \cite{FS_OR2}. The algorithm can be summarized as follows \cite{FS_OR2}:\\
``For each feature $f$ \\
\indent For each value $f_i$ of feature $f$\\
\indent \indent Select set of instances where feature $f$ has a value $f_i$\\
\indent \indent Let $c_j$ be the most frequent class in that set\\
\indent \indent Set the rule: If feature $f$ has value $f_i\; \implies $ class is $c_j$\\
Output the feature/rule with the highest classification accuracy.''\\
Using this algorithm, the classification accuracy of each feature can be calculated and the features can be selected.
\section{Experiment Results \& Discussion}\label{results_dns}
\subsection{Experiment Setup}
\indent\indent MATLAB was used in this work to transform the data from its original state to the new desired dataset representing the previously provided features, perform the feature selection process, and train the corresponding bagging ensemble classification models.
\subsection{Results \& Discussion}
\indent The experiment results are divided into two sections, namely the feature selection results and the bagging ensemble classification model results.
\subsubsection{{Feature Selection}}\mbox{}\\
\indent Tables \ref{feauture_class_correlation} shows the feature ranking using correlation algorithm. Based on this metric, it can be observed in Table \ref{feauture_class_correlation} that the features can be divided into two main subsets. The features within the first subset all have a correlation coefficient above 0.6 while the features in the second subset have a correlation coefficient less than 0.4.
\begin{table}[!htbp]
	\centering
	\caption{Feature Selection Using Correlation}
	\scalebox{0.75}{
		\begin{tabular}{|p{7cm}|p{1.8cm}|}	\hline
			Feature & Correlation \\ \hline
			Number of Unique Characters & 0.663\\ \hline
			Number of Unique Letters& 0.653\\ \hline
			Length of Domain Name& 0.621\\ \hline
			Number of Unique Numbers& 0.329\\ \hline
			Ratio of Numbers to Domain Length& 0.281\\ \hline
			Ratio of Unique Letters to Unique Characters& 0.269\\ \hline
			Ratio of Unique Numbers to Unique Characters& 0.269\\ \hline
			Ratio of Letters to Domain Length& 0.242\\ \hline
		\end{tabular}
	}
	\label{feauture_class_correlation}
\end{table}

\indent Similarly, Table \ref{feauture_class_information gain} shows the information gain of the different features. Again it can be observed that the first subset of features all have an information gain above 0.4 while the second subset has an information gain below 0.15. 
\begin{table}[!htbp]
	\centering
	\caption{Feature Selection Using Information Gain}
	\scalebox{0.75}{
		\begin{tabular}{|p{7cm}|p{1.8cm}|}	\hline
			Feature & Information Gain \\ \hline
			Length of Domain Name& 0.5803\\ \hline
			Number of Unique Characters & 0.4486\\ \hline
			Number of Unique Letters& 0.4220\\ \hline
			Ratio of Letters to Domain Length& 0.1358\\ \hline
			Number of Unique Numbers& 0.1154\\ \hline
			Ratio of Numbers to Domain Length& 0.1096\\ \hline
			Ratio of Unique Letters to Unique Characters& 0.0952\\ \hline
			Ratio of Unique Numbers to Unique Characters& 0.0952\\ \hline
		\end{tabular}
	}
	\label{feauture_class_information gain}
\end{table}

\indent The same observation can be seen in Table \ref{feauture_class_OneR} which shows the class prediction accuracy of the different features. In this case, it is observed that the prediction accuracy of the first subset of features is higher than 80\% while that of the second subset of features is lower than 70\%.
\begin{table}[!htbp]
	\centering
	\caption{Feature Selection Using OneR Classifier}
	\scalebox{0.75}{
		\begin{tabular}{|p{7cm}|p{1.8cm}|}	\hline
			Feature & Accuracy of Rule\\ \hline
			Length of Domain Name& 85.8729\\ \hline
			Number of Unique Letters & 82.0148\\ \hline
			Number of Unique Characters& 81.9193\\ \hline
			Number of Unique Numbers& 68.276\\ \hline
			Ratio of Numbers to Domain Length& 68.0303\\ \hline
			Ratio of Letters to Domain Length& 67.9281\\ \hline
			Ratio of Unique Letters to Unique Characters& 67.5331\\ \hline
			Ratio of Unique Numbers to Unique Characters& 67.5308\\ \hline
		\end{tabular}
	}
	\label{feauture_class_OneR}
\end{table}

\indent These results re-iterate the results shown in \cite{Moubayed_DNS} which illustrated that legitimate domains tend to have more memorable names. In contrast, DGA domains usually have more unique characters with the aim of increasing the randomness of the resulting domain name generated. Accordingly, the subset of features selected as input to the bagging ensemble classification model is made up of 3 features out of the 8 extracted (more than 50\% reduction in the feature size), namely the length of the domain name, the number of unique characters, and the number of unique letters. 
\subsubsection{{Bagging Ensemble Classification Model Performance}}\mbox{}\\ 
\indent As mentioned earlier, a bagging ensemble model was chosen as it can reduce the classification variance of any base model while maintaining its low bias characteristics and improving the classification accuracy \cite{ensemble_variance}. Two different bagging ensemble models are considered, namely a decision-tree bagging ensemble classifier and a $K$-NN bagging ensemble classifier. These base learners where chosen due to their superior performance as illustrated in \cite{Moubayed_DNS}. Similar to our previous work \cite{Moubayed_DNS}, we use accuracy, precision, recall, and F-score as our performance metrics as per the equations in \cite{performance_metrics}. Table \ref{labeled_dataset_results} shows the results of the two bagging ensemble models with the reduced feature set in comparison with the two base learners when the full list of features is used.

\begin{table}[h]
	\centering
	\caption{Performance Evaluation of Classifiers}
	\scalebox{0.75}{
		\begin{tabular}{|p{2.7cm}|p{1.4cm}|p{1.4cm}|p{1.4cm}|p{1.1cm}|}	\hline
			Algorithm & Accuracy (\%) & Precision (\%)& Recall (\%)&F-score\\ \hline
			C4.5 \cite{Moubayed_DNS}&88.1&84.5&95.8&0.89\\ \hline
			K-NN \cite{Moubayed_DNS}&88.2&83.8&94.3&0.89\\ \hline
			Majority-voting Ensemble Classifier \cite{Moubayed_DNS}&88.4&85.5&71.5&0.89\\ \hline
			DT Bagging Ensemble Classifier&87.7&79.2&93.1&0.85\\ \hline
			$K$-NN Bagging Ensemble Classifier&86.7&84.9&80.6&0.82\\ \hline
		\end{tabular}
	}
	\label{labeled_dataset_results}
\end{table}

The results show that both the proposed decision-tree bagging ensemble classifier and $K$-NN bagging ensemble classifier still maintain a high accuracy, precision, and F-score values despite being trained by a significantly smaller feature set. More specifically, we observe that the degradation is at most 1.5\% in terms of accuracy and around 5\% in terms of precision while using less than 50\% of the feature set. This further emphasizes the efficiency of the proposed framework given that it was able to maintain the high accuracy and precision in identifying the malicious/suspicious domains while having a lower computational complexity.
\section{Conclusion \& Future Works}\label{conc_dns}
\indent Domain Name System (DNS) plays in important role in the current IP-based Internet architecture. This is because it performs the domain name to IP resolution. However, the DNS protocol has several security vulnerabilities due to the lack of data integrity and origin authentication within it \cite{DNS_vulnerabilities,DNS_vulnerabilities1}. This work focused on one particular security vulnerability, namely typo-squatting. Typo-squatting refers to the registration of a domain name that is extremely similar to that of an existing popular brand with the goal of redirecting users to malicious/suspicious websites. This is dangerous as it can lead to information threat, corporate secret leakage, and can facilitate fraud. This work extended our previous work in \cite{Moubayed_DNS} and proposed an ensemble-based feature selection and bagging classification model to detect DNS typo-squatting attack. Experimental results illustrated that the proposed framework achieves high accuracy and precision in identifying the malicious/suspicious typo-squatting domains (a loss of at most 1.5\% in accuracy and 5\% in precision when compared to the model that used the complete feature set) while having a lower computational complexity due to the smaller feature set (reduction of more than 50\%).\\
\indent Several potential research directions emerge to extend this work. One potential direction is collecting and exploring the impact of other features such as query sizes and timing. Another direction is studying the impact of a hybrid model that combines multiple techniques such as time series analysis and exploratory data analytics to further our understanding of the data behavior.

\small
\bibliographystyle{IEEEtran}
\bibliography{Ref1}
\end{document}